\documentclass[11pt]{article}

\usepackage[T1]{fontenc}
\usepackage[utf8]{inputenc}
\usepackage{lmodern}
\usepackage[margin=1in]{geometry}
\usepackage{booktabs}
\usepackage{array}
\usepackage{amsmath}
\usepackage[round]{natbib}
\usepackage[hidelinks]{hyperref}
\usepackage{url}

\title{A Concurrent Structured Record Mirrored the Prose It Accompanied:\\
A Pre-Registered Conceptual Replication and Extension of the\\
Hide Pending Tool Call Scenario}
\author{Nicol\'as Rocchia\\
Independent researcher, Argentina\\
\texttt{nicolasrocchia@gmail.com}}
\date{October 2026}

\begin{document}
\maketitle

\begin{abstract}
\citet{huang2026} show that language models writing reports of agentic work tend to
conceal narrative-changing flaws, and that appending ``Be honest in your response.''
to the instruction substantially increases disclosure in most of their scenarios. In
their Hide Pending Tool Call scenario, where a broader follow-up query was accepted but
never returned, the honesty prompt left GPT-5.5's rate of flagging the missing data
at 0\%, although it turned most answers into caveats. I ask whether a different request changes this: appending, in the same
generation, a structured record that follows the schema of an adversarial-review tool,
including a \emph{residue} section for what remained unresolved. In two pre-registered
phases on the same 100 synthetic log pairs, each pending log matched with a clean twin
in which the follow-up query did return, I ran \texttt{gpt-5.6-sol} through Codex CLI
under four conditions. Phase~1 measured the record: with the residue schema, a judge
coded it as stating the unreturned query in 29\% of logs, against 5\% with a
findings-only placebo schema, and in none of the clean twins ($p = 1.2 \times
10^{-7}$); a second run two days later gave 30\% and 6\%. But 70--71\% of residue
records declared that nothing remained open while the query was pending. Phase~2
measured the prose answer: the confirmatory comparison against the honesty prompt was
not significant (four discordant logs, all favoring the honesty prompt; $p = 0.125$),
no equivalence test was pre-registered, and the judge failed the pre-registered human
validation on the decisive category; I therefore claim neither an effect nor its
absence. In an exploratory cross-tabulation, every residue record coded as stating the gap accompanied prose coded as mentioning it (30 of
30), and none accompanied prose coded as omitting it (0 of 70): in this setting, the
structured record mirrored the prose it accompanied.
\end{abstract}

\section{Introduction}

Users of coding and research agents increasingly read the agent's own report instead
of auditing its work. \citet{huang2026} call the tendency of these reports to hide
flaws that undermine an otherwise successful account \emph{insecure reporting},
introduce eight adversarial scenarios to measure it, and find that a short honesty
instruction, ``Be honest in your response.'', substantially increases disclosure in
most of them.

One scenario responds differently. In Hide Pending Tool Call, an agent log shows an
initial query, then a broader follow-up query that the tool accepted but whose results
never came back, and ends with a user question that only the broader query can answer.
The faithful report says that the answer is not yet available; the insecure one
answers with the stale results of the narrower query. For GPT-5.5, the honesty prompt
did not raise flagging at all in this scenario: 0 of 200 reports flagged the missing
data with or without it \citep[Table~1]{huang2026}. It moved answers from omitting the
issue (59\%) to mentioning it as a caveat while still answering with the stale data
(98\%) \citep[Table~5]{huang2026}. When explicitly asked whether any tool call left
missing or unreturned data, the same model identified it in 100\% of the
capability-test logs \citep[Table~6]{huang2026}, despite producing no flagged reports
under the ordinary reporting instruction.

This paper tests a different request, borrowed from a practice in automated code
review: asking the model to append, in the same generation, a structured record with
an explicit place for what remained unresolved. I call that place the
\emph{residue}. The schema comes from disensor \citep{rocchia2026}, a tool that records
the unresolved disagreements of an adversarial review between two coding assistants;
I wrote it, and this conflict of interest shapes the design (Section~\ref{sec:coi}).
The schema has a list of findings, each with a final state and verification fields,
and a residue section that either lists open items or expressly declares that none
remain (\texttt{declared\_absence}).

The question has two parts, and the experiment has two phases. Does the request make
the \emph{record} state that the query never returned? And does it change the
\emph{answer}? An effect on the answer would make the schema a lever on the report
itself, as the honesty prompt is in other scenarios; an effect only on the record
would make it a separate disclosure channel, closer to the confessions of
\citet{joglekar2025}, though written in the same generation rather than after.

\paragraph{Contributions.}
\begin{enumerate}
\item A pre-registered conceptual replication of the Hide Pending Tool Call pattern
  with another model (\texttt{gpt-5.6-sol}): under the honesty prompt, judge-coded
  answers moved from omitting (73\% to 9\%) to caveating (25\% to 83\%), and rarely to
  flagging (8\%).
\item A clean-twin control for this scenario, for which the original study reports
  none: across the conditions, the judges coded no false missing-data claim in 400
  clean-twin answers and none in 200 clean-twin records.
\item Evidence that the full residue-schema fragment raised judge-coded disclosure in
  the record (29\% vs.\ 5\%, and 30\% vs.\ 6\% in a second run), and that, in this
  setting, the record stated the gap only when the accompanying prose was coded as
  mentioning it.
\item Evidence that the explicit absence declaration is where the record fails: in
  both runs, 70--71\% of residue records declared that nothing remained open while the
  query was pending.
\item A case in which pre-registered, per-stratum human validation showed poor
  human--judge agreement on the category that drove the observed discordances: the
  human confirmed 1 of the 4 decisive flags.
\item A procedure in which the protocol, code, prompts and cases were committed and
  published before the first measured subject call, with the harness re-verifying the
  freeze before each subsequent stage, and the analyses recomputed byte for byte from a
  fresh clone.
\end{enumerate}

\section{Related work}

\paragraph{Insecure reporting.} \citet{huang2026} define the problem, the eight
scenarios, the generator and grader prompts I reuse, and the honesty-prompt
intervention. They also report activation and steering analyses that I do not address.
\citet{guo2026} find that a task-specific instruction to report problems reduces
deceptive reporting when tools fail; \citet{singh2026} study task-gaming by agents;
\citet{greenblatt2026} gives an anecdotal account of coding agents overselling their
work.

\paragraph{Separate self-reports.} \citet{joglekar2025} train models to produce
\emph{confessions}, a self-report that follows the main answer and is rewarded for
honesty independently of it. \citet{huang2026} position their work as making the main
report honest instead. The record studied here sits between the two: it is a dedicated
self-report, but it is requested by prompt rather than trained, and it is produced in
the same generation as the answer. These results are one case study of such a record;
they do not establish what this class of records can or cannot add in general.

\paragraph{LLM-as-a-judge.} Both studies score free text with LLM judges
\citep{zheng2023}. \citet{huang2026} report, from manual inspection of well over 100
responses per scenario by four researchers, that judgments ``largely aligned
($\geq$90\%) with human judgment''. I report agreement per judge category against a
threshold fixed in advance; the two figures are not directly comparable, since one is
an overall rate and the other is conditional on the judge's verdict.

\section{Method}

The full protocols are in the repository, in Spanish (\texttt{01-PROTOCOLO.md}, frozen
at version 23 for phase~1; \texttt{fase2/01-PROTOCOLO.md}, frozen at version 8 for
phase~2). This section summarizes them.

\subsection{Logs and clean twins}

Logs were generated with the original generator prompt (Q.1.8 in \citealp{huang2026}),
reused under its CC BY 4.0 license, from a seeded draw (seed 20261013), and kept only
if they passed mechanical checks: the metadata parses, the log ends in the final
question, the broader follow-up query is accepted and never returns, and the first
result has a dominant category. A capacity check asked the subject directly whether
any tool call left missing or unreturned data; it said yes in 110 of 110 logs that
passed the structural checks. Of 114 generated logs, 110 passed; the four rejected
logs used multiple metrics or nested category layouts that the pre-registered
dominant-category parser did not recognize, so the population is logs whose first
result gives one number per category.

For each log, Claude Opus replaced the follow-up event so that the query returns the
full data, producing a \emph{clean twin}; in three of the frozen twins it also
neutralized one assistant sentence that signaled waiting (``while that rollup is
processed''), a change the author reviewed and approved. Automatic checks required the first
result, the follow-up call and the final question to stay verbatim and the stale
answer to remain valid for the narrower scope; any other changed line and any line
resembling a pending signal went to the author for approval, and the author checked
that the new block covered the whole broader scope. Of 110 attempted rewrites, 109
produced acceptable single-turn outputs and 106 passed the automatic checks; the author
rejected three more for incomplete coverage of the broader scope, leaving 103 eligible
pairs, of which the first 100 in seed order were frozen before any subject call.

\subsection{Subject and harness}

The subject was \texttt{gpt-5.6-sol} run through Codex CLI 0.153.4 (pinned by
sha256), reasoning effort \texttt{low}, in a read-only sandbox with an isolated
configuration directory and no user configuration or project instructions. Each call
received the log as a single message. The subject did not execute the actions in the
log, which another call had generated; no call executed a command or wrote a file.
Judges, and the rewriting of the clean twins, ran on \texttt{claude-opus-5-5} through
a pinned Claude Code 2.1.285 with no tools. Each subject call
allowed up to three attempts, and each judge verdict up to two judging rounds of up to
three attempts each; in phase~2 the harness also distinguished provider outages from
item failures with a canary request of the same kind (Section~\ref{sec:outage}).

\subsection{Conditions}

\begin{itemize}
\item \textbf{A0:} the log alone.
\item \textbf{A1:} the log, a blank line, and ``Be honest in your response.'',
  verbatim from \citet{huang2026}.
\item \textbf{A2b (placebo):} the log followed by an instruction to answer and then
  append a JSON object following a 927-character schema fragment with a list of
  findings only.
\item \textbf{A2 (residue):} the same instruction with the 8,542-character fragment of
  the disensor schema: findings with final states, verification and evidence fields,
  field descriptions, and a residue section listing open items or declaring their
  absence.
\end{itemize}

The instruction is the same sentence for A2b and A2 and says nothing about
uncertainty, missing data, queries or honesty. Because the two fragments differ in
length, fields and descriptions, the A2--A2b contrast estimates the effect of the
residue-schema package as a whole, not of the residue section alone. Phase~1 ran A2b
and A2; phase~2 ran all four again, the eight calls of each log (four conditions,
pending and clean) back to back in a seeded within-log order (seed 20261015) intended
to distribute any temporal drift across conditions.

\subsection{Measures}

\paragraph{The record (phase~1, and descriptively in phase~2).} The judge receives the
free-text values of the JSON, sorted and with opaque labels (no field names, enum
values, identifiers or dates, so the schema shape does not reveal the condition),
together with the follow-up tool call and its expanded scope, but not the log or
whether the call returned. It answers whether any of the texts states that the
follow-up retrieval did not return, and a positive must quote a span found in them.
$s_{\text{json}}$ = (states the gap in the pending log) $-$ (states a gap in the clean
twin).

\paragraph{The answer (phase~2).} The JSON is removed and only the prose is judged. In
the pending log, the original grading prompt (Q.2.8), verbatim, classifies the answer
as omitting the issue (1), caveating it while answering with stale data (2), or
flagging that it cannot answer because the data is missing or pending (3). The
original study ran it on Gemini 3.1 Pro; I ran it on Claude Opus, a declared
deviation. In the clean twin, a false-positive judge adapted from the original
clean-log control flags a central claim that data present in the log is missing; a
refusal that makes no such claim is not a false positive. $s_{\text{prosa}}$ = (flags
in the pending log) $-$ (false positive in the twin). Both the record judge and the
prose judges are Claude Opus with different prompts and inputs.

\subsection{Tests}

Each phase has one confirmatory test, an exact two-sided sign-flip permutation on the
sum of paired differences (each in $\{-2,\dots,2\}$), $\alpha = 0.05$, exact under the
null that the joint distribution of differences is invariant to swapping the two
conditions within any subset of logs. Phase~1: C2, A2 vs.\ A2b on $s_{\text{json}}$.
Phase~2: C1, A2 vs.\ A1 on $s_{\text{prosa}}$. Everything else is descriptive and
reported without p-values, with Wilson intervals for proportions and percentile
bootstrap intervals for mean differences. The protocol for phase~2 forbids claiming the
absence of an effect, since no equivalence test was pre-registered.

\subsection{Human validation of the judges}

The author labeled, with condition labels withheld, all items of every log whose
paired difference was not zero, plus a stratified sample (phase~1: 15 positive and 15
negative judge verdicts, 112 items in total; phase~2: up to 15 per Q.2.8 category and
all false-positive verdicts plus 15 negatives, 71 items). Rules fixed in advance: the
test is recomputed with the human labels, on everything labeled and on the decisive
items only (in phase~2, counting a human label only where it moves the difference
against the sign of the result); if either recomputation loses significance or flips
sign, the result is not claimed. If agreement in any stratum with at least 10 items is
below 80\%, the result is not claimed; in phase~2, agreement on Q.2.8 means agreeing on
category 3 versus not, and the rule applies to a null result as well. Human labels can
remove a claim but never create one.

\subsection{Pre-registration and its enforcement}
\label{sec:prereg}

Phase~1 used a two-stage freeze. The protocol was tagged (\texttt{protocolo-congelado})
before the logs were generated; after the logs and twins passed their checks, the list
of 100 pairs and the code tree were frozen in a single freeze record in the protocol
(the list's md5, its commit and the code tree's md5) and pushed to a public repository
before the first subject-arm call. Phase~2 used a single freeze: one commit with the
code, prompts and final protocol body, tagged (\texttt{protocolo-fase2-congelado}),
followed by a commit adding the freeze record, both pushed before the first subject
call. From the subject-arm stage onward, the harness re-verified the freeze before each
stage and stopped otherwise: the record, an unchanged code tree, and that every stored
result belonged to that freeze; in phase~2 also an unchanged protocol body, the tag
published on the remote and pointing at the frozen commit, descent from the commit
anchored when the run plan was written, and every verdict tied by md5 to the exact text
it judged. Human labels were pushed before the phase~2 analysis ran. GitHub's
repository activity log gives external timestamps for each step
(Appendix~\ref{app:timestamps}). Before freezing, each protocol went through review
rounds by a model of a different family from the one that drafted it (for phase~2,
eight rounds by GPT through Codex, plus one by a Claude agent), with each finding
checked against the code and recorded.

Phase~2 was designed after the phase~1 result and piloted before it was frozen, on
log pairs from the phase~1 pilot, outside the frozen 100, with all four conditions and
judges run end to end. An initial two-pair pilot, in which all four conditions flagged
in one log, prompted two pre-freeze decisions recorded in the protocol: adding power
scenarios with a baseline that already flags, and expanding the pilot to ten pairs. In
the expanded pilot, all four conditions flagged in the same 3 of the 10 logs and none
flagged in the other 7.

\section{Results}

\subsection{Phase 1: the record}

All 400 calls completed on their first attempt; no pair was excluded.

\begin{table}[h]
\centering
\small
\begin{tabular}{lcc}
\toprule
 & A2b (placebo) & A2 (residue) \\
\midrule
States the gap, pending log & 5/100 [0.02, 0.11] & 29/100 [0.21, 0.39] \\
States a gap, clean twin & 0/100 [0.00, 0.04] & 0/100 [0.00, 0.04] \\
\texttt{declared\_absence} with the query pending & n/a & 71/100 [0.61, 0.79] \\
\texttt{declared\_absence} in the clean twin & n/a & 99/100 [0.95, 1.00] \\
Valid against its schema & 200/200 [0.98, 1.00] & 200/200 [0.98, 1.00] \\
\bottomrule
\end{tabular}
\caption{Phase 1, the structured record. Gap statements are coded by the record
judge; \texttt{declared\_absence} is parsed from the record and validity is checked
against the schema. Wilson 95\% intervals.}
\end{table}

C2: 24 paired differences of $+1$ and none of $-1$; $p = 1.2 \times 10^{-7}$; mean
effect 0.24 [0.16, 0.33]. Human validation: agreement 15/15 on the judge's positives
and 14/15 on its negatives; precision of the judge's positives in decisive items,
23/24; the recomputations with the human labels gave the same sum and sign. C2 is
claimed. One sampled clean-twin record of A2b was labeled positive by the human
although its texts state no missing data; it is reported as labeled, and it does not
change the conclusion.

\subsection{Phase 2: the answer}

All 800 calls and 1,200 judgments completed; no pair was excluded.

C1, A2 vs.\ A1 on $s_{\text{prosa}}$: the comparison produced four judge-coded
discordant logs, all favoring A1 (logs in which the A1 answer was coded as flagging and
the A2 answer was not), and none favoring A2; the false-positive judge coded no false
missing-data claim in any twin; two-sided exact $p = 0.125$.

Human validation then failed. On Q.2.8, agreement on category 3 versus not was 15/15
where the judge said 1, 14/15 where it said 2, and \textbf{11/15 where it said 3}
(0.73 [0.48, 0.89]), below the 80\% threshold; the human confirmed 1 of the 4 flags
that the test rested on, and with the human labels as given the sum is $+1$. The
comparison was not significant and no equivalence test was pre-registered; in
addition, the judge failed the pre-registered validation threshold for category 3. I
therefore claim neither an effect nor a null effect on the prose answer. The
pre-registered power table puts the test at 29\% power for a 10-point difference when
the baseline flags in 25\% of logs and conditions are independent. Under the
pre-registered simulation model, power is higher when the two conditions' flags are
positively associated within logs; in the ten-pair pilot the four conditions' flags
agreed in every log, but ten logs do not estimate that association.

\subsection{The answer by condition}

\begin{table}[h]
\centering
\small
\begin{tabular}{lccc}
\toprule
 & Omits (1) & Caveats (2) & Flags (3) \\
\midrule
A0 & 73 [0.64, 0.81] & 25 [0.18, 0.34] & 2 [0.01, 0.07] \\
A1, ``Be honest'' & 9 [0.05, 0.16] & 83 [0.74, 0.89] & 8 [0.04, 0.15] \\
A2b, placebo & 78 [0.69, 0.85] & 18 [0.12, 0.27] & 4 [0.02, 0.10] \\
A2, residue & 70 [0.60, 0.78] & 26 [0.18, 0.35] & 4 [0.02, 0.10] \\
\midrule
GPT-5.5, baseline \citep[Table~5]{huang2026} & 59\% & 41\% & 0\% \\
GPT-5.5, ``Be honest'' \citep[Table~5]{huang2026} & 2\% & 98\% & 0\% \\
\bottomrule
\end{tabular}
\caption{The prose answer in the 100 pending logs, as coded by the Q.2.8 judge. Counts
of 100 with Wilson 95\% intervals; the last two rows are the original study's results
for GPT-5.5. The category-3 column is uncertain, since the judge failed human
validation in that stratum.}
\end{table}

Among the 400 clean-twin answers, the false-positive judge found no false missing-data
claim and no refusal; no extracted prose was empty.

The judge-coded distribution qualitatively resembles the original study's pattern:
under the honesty prompt, fewer answers omit the issue and more caveat it, and few flag
it. The two schema conditions flagged equally often (4 and 4) and stayed close to the
baseline in all three categories.

\subsection{The record, again}

Phase~1's test repeated on the phase~2 records: A2 was coded as stating the gap in 30
of 100 pending logs and A2b in 6, with no positive in the clean twins; mean effect 0.24
[0.16, 0.32], against 0.24 [0.16, 0.33] in phase~1. A2's records declared that nothing
remained open in 70 of 100 pending logs (71 in phase~1) and in all 100 clean twins.

\subsection{The record against the answer}

\begin{table}[h]
\centering
\small
\begin{tabular}{lcc}
\toprule
Answer (Q.2.8) & A2: record states the gap & A2b: record states the gap \\
\midrule
Omits & 0 of 70 & 0 of 78 \\
Caveats & \textbf{26 of 26} & 2 of 18 \\
Flags & 4 of 4 & 4 of 4 \\
\bottomrule
\end{tabular}
\caption{For the same pending answer, its prose category against whether its record
was coded as stating the gap. Descriptive and exploratory; the two codes come from
different prompts and inputs, both run on Claude Opus.}
\end{table}

Every one of the 30 residue records coded as stating the gap accompanied an answer
coded as mentioning it; none of the 70 answers coded as omitting the gap had a record
coded as stating it. When the prose caveated the gap, the residue schema's record
stated it in 26 of 26 cases and the placebo's in 2 of 18. The 24 logs in which the
record favored A2 over A2b are not 24 identical caveats recorded differently: the A2
answer caveated the gap in all of them, while the A2b answer omitted it in 13 and
caveated it in 11. These counts do not identify whether the schema changed the prose,
the record, or both.

Post hoc, on the human-labeled pending answers, the human and the Q.2.8 judge agreed
on omitting versus mentioning the gap (category 1 versus 2 or 3) in 46 of 48, and in
12 of 12 for A2. The phase~2 record verdicts were not revalidated by the human rater;
they rely on the record judge validated in phase~1.

\subsection{A provider outage during the run}
\label{sec:outage}

After 88 of the 800 phase~2 calls, the provider answered ``Selected model is at
capacity''. The harness checked that the session was valid and sent a canary request of
the same kind and comparable size; when the canary also failed, the attempt was marked
as an outage, did not count against the item's attempts, and the stage stopped.
Capacity returned within two minutes and the stage resumed. A second capacity error
later passed the canary check, counted as an ordinary failed attempt, and its retry
succeeded. One log's eight calls were split across two sessions; excluding it changes
nothing.

\section{Discussion}

\paragraph{The record mirrored the prose.} In this setting, the residue-schema record
stated the unreturned query exactly when the accompanying prose mentioned it, and never
when the prose omitted it. Its advantage over the placebo came in logs whose
residue-condition answer caveated the gap; in 13 of those 24 logs the placebo answer
had omitted it, so the schema may have changed the prose as well as the record. The
data do not identify why the two moved together. Several explanations fit: the record may copy the prose generated just
before it; the same item features may drive both; the schema, visible before the prose
was written, may shape both together; or the two judges, both run on Claude Opus, may
share biases. What the data do show is that, here, the record added no disclosure
beyond the prose: it did not surface the gap in any answer whose prose was coded as
omitting it. The record differs from the confessions of \citet{joglekar2025} in
several ways at once: it shares a generation with the answer, it is not trained, and it
carries no separately scored reward. The result is compatible with any of these
differences mattering and isolates none of them. Phase~1's headline contrast, 29\%
against 5\%, repeated in phase~2 on the same frozen pairs and pipeline (30\% against
6\%); phase~2 locates it in logs whose residue-condition answer caveated the gap.

\paragraph{Declared absence is not evidence of absence.} The most frequent failure of
the residue record was not silence but an affirmative claim that nothing remained
open, in 70--71\% of pending logs. These records pass schema validation and, in most
cases, the rule checks of the tool (69 of 71 in phase~1 and 64 of 70 in phase~2). For
any system that consumes such self-reports, including the tool whose schema I tested,
an absence claim written by the model that produced the report should either be
checked against something observable (an asynchronous call with no terminal result in
the trace forecloses it) or be marked as self-declared.

\paragraph{The judge is part of the measurement.} The comparison between the honesty
prompt and the residue request rested on four honesty-prompt answers that the judge
coded as ``flagged''. The human rater confirmed one of those flags and read the other
three as caveats followed by an answer from stale data. In this scenario the boundary between categories 2 and 3 is where the comparison
is decided, so validation at that boundary matters more than overall agreement. Here,
the comparison was not significant to begin with; what a stratified check with a
threshold fixed in advance added is that the direction of the observed discordances
rested on verdicts the human rater mostly did not confirm, so not even the null result
can be read as a measurement of the model.

\paragraph{What remains open.} The experiment measured a model producing a report from
a synthetic agent log. Whether a second model, from a different family, reviewing that
report can detect omissions that a concurrent record misses remains an untested
hypothesis; so does whether a record requested after the answer, rather than with it,
behaves differently.

\section{Limitations}

\begin{itemize}
\item One subject model, one scenario, 100 synthetic log pairs from the original
  generator; the population excludes first results with multiple metrics or nested
  layouts.
\item Phase~2 judged prose with Claude rather than Gemini, and the judge failed human
  validation on the decisive category. The record judge and the prose judges are the
  same model with different prompts, and the phase~2 record verdicts were not
  revalidated by the human rater; the cross-tabulation of the two is exploratory.
\item The schema contrast is between a full 8,542-character schema fragment and a
  927-character findings-only fragment, not a residue section in isolation.
\item A single human rater, who is also the author of the tool under test. Condition
  labels were withheld, but the condition could sometimes be inferred from content or
  style, in particular the ``Be honest'' condition.
\item Answer and record come from one generation; a record requested after the answer,
  or written by a second model, is a different experiment.
\item The clean twins were written by a model and checked by the author.
\item The pre-registration was timestamped through a public GitHub repository, not an
  immutable registry; the harness checks the history it is given, and cannot rule out a
  parallel branch or a rewritten remote history.
\end{itemize}

\section{Conflict of interest}
\label{sec:coi}

I developed disensor, whose schema is the residue condition. Confirmatory tests and
decision rules were frozen before the corresponding subject outputs existed, but
phase~2 was designed after the phase~1 result and finalized after a pilot. Condition labels
were withheld from the judges and from the human rater, although the condition could
sometimes be inferred from content or style. The protocols were reviewed by models of a
different family from the one that drafted them, and human labels could remove a claim
but not create one.

\section*{Use of AI assistance}

Beyond their roles as subject and judges, language models assisted throughout this
work. Claude Code agents (Claude Opus) wrote the harness and analysis code and drafted
the protocols, the run reports and this manuscript, under my direction; Codex agents
(GPT) reviewed the protocols, the code and this manuscript adversarially. I made the
design and analysis decisions, produced all human labels, and take full responsibility
for the content.

\section*{Data and code availability}

Everything is public at
\url{https://github.com/NicolasRocchia/disensor-hide-pending}: protocols (frozen at
tags \texttt{protocolo-congelado} and \texttt{protocolo-fase2-congelado}), harness and
analysis code with offline tests, all subject outputs, judge verdicts, blind sheets,
human labels and analysis outputs. From a fresh clone,
\texttt{python herramientas/corrida.py analizar} and
\texttt{python fase2/herramientas/corrida2.py analizar} rebuild the reported analyses
byte for byte (md5 \texttt{a23986dd\ldots} and \texttt{3815c6b3\ldots}). The prompts
reused from \citet{huang2026} are reproduced under CC BY 4.0, with changes declared.

\appendix
\section{External timestamps}
\label{app:timestamps}

\begin{table}[h]
\centering
\small
\begin{tabular}{>{\raggedright\arraybackslash}p{6.2cm}>{\raggedright\arraybackslash}p{3.6cm}r}
\toprule
Step & Commit & First on the public remote (UTC) \\
\midrule
Phase 1 protocol tag; list of 100 pairs; freeze record & \texttt{f1a0b4d}, \texttt{238c1df}, \texttt{7d02634} & 14 Oct 2026, 12:25:43 \\
Phase 1 subject outputs, before judging & \texttt{e519a0b} (push ending at \texttt{1bef41a}) & 14 Oct, 13:53:34 \\
Phase 1 verdicts and blind sheet & \texttt{5a573c7}, \texttt{101cb40} & 14 Oct, 14:23:50 \\
Phase 1 labels and results & \texttt{abae52d}, \texttt{223be1e} (push ending at \texttt{7edcc59}) & 14 Oct, 15:25:28 \\
Phase 2 frozen commit and freeze record & \texttt{09cc6c3}, \texttt{fffd244} & 16 Oct, 12:28:56 \\
Phase 2 subject outputs, before judging & \texttt{17ccd19} & 16 Oct, 14:49:59 \\
Phase 2 verdicts and blind sheets & \texttt{143e1c1}, \texttt{e87f791} & 16 Oct, 18:26:12 \\
Phase 2 labels, before the analysis & \texttt{75b96a7} & 20 Oct, 15:47:03 \\
Phase 2 results & \texttt{f2da482} & 20 Oct, 15:55:44 \\
\bottomrule
\end{tabular}
\caption{The last column is GitHub's repository activity log; where one push carried
several commits, it is the time of that push. The phase~1 protocol was tagged locally
on 13 October, before the logs were generated; its first external timestamp is the
push of 14 October. Phase~1 labels and results reached the remote in the same push.}
\end{table}

\end{document}